\documentclass[aps,prd,onecolumn,nofootinbib]{revtex4-2}

\usepackage{amsmath,amssymb,amsfonts}
\usepackage{graphicx}
\usepackage{hyperref}
\usepackage{xcolor}
\usepackage{bm}
\usepackage{physics}
\usepackage{float}
\usepackage{graphicx}
\usepackage{subcaption}
\usepackage[utf8]{inputenc}
\usepackage{textgreek}

\hypersetup{
    colorlinks=true,
    linkcolor=blue,
    citecolor=blue,
    urlcolor=blue
}

\begin{document}

\title{Scalar-Field Reconstruction of Ricci--Gauss--Bonnet Dark Energy in Ho\v{r}ava--Lifshitz Cosmology}

\author{Surajit Chattopadhyay$^{1}$}
\email{surajitchatto@outlook.com, schattopadhyay1@kol.amity.edu}
\affiliation{$^{1}$Department of Mathematics, Amity University Kolkata, \\
Major Arterial Road, Action Area II, Newtown, Kolkata 700135, India}

\date{\today}

\begin{abstract}
This paper reports a Ricci-Gauss-Bonnet (RGB) dark energy model within the framework of Hořava-Lifshitz cosmology and presents a scalar-field reconstruction of the effective dark energy sector. In a spatially flat FRW background with a power-law scale factor, we derive analytical expressions for cosmological parameters, scalar field kinetic term, and the reconstructed potential. The reconstructed EoS parameter exhibits smooth transition toward a cosmological-constant-like regime at late times for suitable choices of the model parameters. The classical stability of the model is analyzed through the squared sound speed, and stable regions of the parameter space are identified. Finally, the generalized second law of thermodynamics is investigated at the apparent horizon, and it is shown that the total entropy variation remains non-negative in this model. From these results it can be concluded that the model provides a theoretically consistent description of late-time acceleration, with physical viability maintained within a specific range of the model parameters.
\end{abstract}

\maketitle

\section{Introduction}

\textcolor{black}{A major milestone in modern cosmology was the discovery of the late-time accelerated expansion of the Universe \cite{Perlmutter1999,Riess1998}. Observations of Type Ia supernovae by two independent collaborations showed that the cosmic expansion is accelerating rather than decelerating \cite{Perlmutter1999,Riess1998}. The cosmological origin of the observed redshifts was further confirmed through time dilation in supernova spectral evolution \cite{Foley2005}. In addition, ultraviolet and optical observations revealed the intrinsic diversity of Type Ia supernovae and highlighted the importance of controlling systematic effects in precision cosmology \cite{Wang2012}. These observations provided the first direct evidence of cosmic acceleration and motivated the development of dark energy models and modified gravity theories. Subsequent observations have firmly established that the late-time Universe is undergoing accelerated expansion, commonly attributed to an exotic component with negative pressure known as dark energy \cite{DES2016,Copeland2006,Frieman2008,Peebles2003,Li2013,Yoo2012,Sahni2006,Oks2021}.} 

\textcolor{black}{Large-scale surveys, particularly the Dark Energy Survey, have improved our understanding of dark energy beyond the cosmological constant by providing precise data on the expansion history and growth of cosmic structures \cite{DES2016}. To explain the accelerated expansion of the universe, several models have been proposed in the literature \cite{Copeland2006,Peebles2003,Li2013,Yoo2012,Sahni2006}. Review articles discuss their successes and remaining challenges and highlight the need for models consistent with observations \cite{Frieman2008,Oks2021}. Alfano et al. \cite{Alfano2025} have shown that although the Dark Energy Spectroscopic Instrument (DESI) observations do not rule out dynamical dark energy, the $\Lambda$CDM model remains statistically preferred because it involves the fewest free parameters. Using gamma-ray burst correlations as high-redshift distance indicators to test several cosmological models, the analysis of \cite{Alfano2025} found no strong model preference overall, though including a particular high-redshift data point slightly disfavors the concordance ΛCDM scenario. These developments have renewed interest in alternative gravitational frameworks capable of describing evolving dark energy, including scenarios based on Ho\v{r}ava--Lifshitz gravity and related modified gravity models \cite{Dunsby2016,Luongo2016arXiv,Dunsby2024}.~}

Modified gravity has emerged as a compelling theoretical framework to explain cosmic acceleration without invoking an explicit dark energy component, offering rich phenomenology across inflationary, bouncing, and late-time cosmological regimes \cite{Koyama2016,Joyce2016,Clifton2012,Myrzakulov2013,Petrov2023,Nojiri2017,Bamba2015}. Extensive theoretical developments and cosmological tests have demonstrated that deviations from General Relativity can naturally account for large-scale observations while remaining consistent with local gravity constraints \cite{Koyama2016,Clifton2012}. Comparative studies have further clarified the conceptual and observational distinctions between dark energy models and modified gravity scenarios, emphasizing the role of gravitational degrees of freedom in driving cosmic acceleration \cite{Joyce2016}. Recent reviews continue to consolidate these advances, highlighting modified gravity as a unifying framework capable of addressing early- and late-universe dynamics within a single theoretical setting \cite{Nojiri2017,Bamba2015,Petrov2023}.

\textcolor{black}{Recent works in modified gravity and generalized entropy have offered additional insight into the dynamical behaviour of the Universe. Bouncing cosmological scenarios in Aether scalar–tensor theory have been studied to understand the avoidance of cosmological singularities and the stability of the early Universe \cite{Rani2026}. Tsallis entropy correction has been used to study the stability of dark-energy dominated universes and the role of generalized entropy in cosmological dynamics \cite{Alruwaili2026}. Modified gravity models with non–zero torsion have also been analysed from the observational point of view using cosmographic parameters together with thermodynamical considerations \cite{Azhar2025}. These works show that torsion-based gravity frameworks can explain the late-time accelerated expansion of the Universe while remaining consistent with observational data and thermodynamic conditions \cite{Azhar2025}. Similar investigations have also been carried out in higher–dimensional gravity, where the cosmological viability of five–dimensional Einstein–Chern–Simons gravity has been studied using cosmographic and thermodynamic analysis \cite{Azhar2025EChS}. Nonlinear interactions in modified gravity have also been studied and are found to influence the phase portraits and dynamical stability of cosmological models \cite{Usman2025}.}

\textcolor{black}{Holographic dark energy models based on the holographic principle have received much attention in recent years \cite{Dubey2021,Wang2023,DelCampo2011,Wang2005,Ma2009,Singh2020}. The original idea has been extended by introducing generalized infrared cutoffs constructed from geometric quantities. In this direction, \cite{NojiriOdintsov2017} proposed a covariant generalized holographic dark energy model where the infrared cutoff depends on curvature invariants and cosmological variables \cite{NojiriOdintsov2017}. Earlier studies showed that holographic vacuum fluctuations can produce dynamical dark energy behaviour, including phantom phases and crossing of the cosmological-constant boundary \cite{ElizaldeNojiriOdintsovWang2005, NojiriOdintsov2006}. Later, \cite{NojiriOdintsovPaul2022} showed that generalized holographic models based on modified entropy relations can describe both early and late-time cosmic acceleration \cite{NojiriOdintsovPaul2022}. These works motivate the curvature-based Ricci--Gauss--Bonnet dark energy used in this work.}

\textcolor{black}{Curvature-based dark energy models involving the Ricci scalar and Gauss–Bonnet invariant have been proposed to capture higher-order gravitational effects \cite{Dabash2025,Bueno2019,AguilarGutierrez2023}. In parallel, Ho řava-Lifshitz gravity introduces anisotropic scaling between space and time and provides a renormalizable theory of power-counting with modified cosmological dynamics at high energies \cite{Mukohyama2010,Horava2009,Bajardi2021,Calcagni2009,Li2014,Czuchry2023}. The theory has also been applied to important cosmological issues such as the baryon-to-entropy ratio, suggesting that modified Ho\v{r}ava--Lifshitz gravity can describe the thermal history of the Universe \cite{Jawad2022a,Jawad2023,Jawad2022b,Chaudhary2024}. Recent observational analyses have also placed tighter bounds on the free parameters of Ho\v{r}ava--Lifshitz gravity, strengthening the phenomenological relevance of these models \cite{Jawad2022a,Jawad2023,Jawad2022b,Chaudhary2024}. The current work is inspired by \cite{Jawad2022b}, which investigated the cosmological implications of a Barrow holographic dark energy model within the framework of deformed Hořava–Lifshitz gravity, exploring how quantum-gravity-inspired entropy corrections influence dark energy dynamics and cosmic expansion history.} 

\textcolor{black}{The above discussion should also be understood in the wider context of the modified gravitational dynamics present in Ho\v{r}ava--Lifshitz cosmology. In this framework, Lorentz invariance is broken at high energies and higher--order spatial curvature terms appear in the gravitational action. Because of these features, the cosmological background evolution can differ from that predicted by standard Einstein gravity. As a result, the effective dark--energy sector reconstructed in the present work reflects not only the curvature contributions associated with the Ricci scalar and the Gauss--Bonnet invariant, but also the modified gravitational structure of the HL theory. These modifications may influence the dynamical behaviour of the reconstructed scalar field and the corresponding stability and thermodynamical properties of the cosmological model. At the same time it is important to note that Ho\v{r}ava--Lifshitz gravity is known to face certain theoretical challenges, such as issues related to the extra scalar graviton mode, strong coupling behaviour, and consistency of the infrared limit, which have been discussed in the literature. \textcolor{black}{In this context, it is worth noting that the minimal Ho\v{r}ava--Lifshitz cosmological model may face certain inconsistencies. Luongo et al.~\cite{Luongo2016arXiv} demonstrated that, based on cosmographic constraints and statistical tests, the model is kinematically and statistically disfavoured when compared with standard $\Lambda$CDM and other dark energy scenarios.}~Although a detailed investigation of these aspects is beyond the scope of the present work, we briefly mention here these limitations, and focus our analysis on the phenomenological cosmological implications of the HL framework.}

\textcolor{black}{In the present work we study a Ricci--Gauss--Bonnet (RGB) dark energy model in the framework of Ho\v{r}ava--Lifshitz cosmology. The RGB model combines the Ricci scalar and the Gauss--Bonnet invariant, allowing both local curvature effects and higher-order geometric contributions to influence cosmic evolution \cite{Saridakis2018,Iqbal2018,Ahmed2020,Rudra2023,Pradhan2025}. In the framework of Ho\v{r}ava--Lifshitz cosmology the Lorentz symmetry is broken at high energies and higher-order spatial curvature terms. In this context, with a unified treatment of early- and late-time cosmic acceleration under observational constraints, the Ricci--Gauss--Bonnet dark energy model is particularly well motivated because it complements the ultraviolet structure of the theory  \cite{Dubey2025}. Therefore, investigating curvature-based dark energy models within Hořava–Lifshitz cosmology can provide additional insight into the dynamics of late-time cosmic acceleration.}

A closely related study of Gauss--Bonnet dark energy in Ho\v{r}ava--Lifshitz cosmology is Ref.~\cite{LepeOtalora2018}. In that work~\cite{LepeOtalora2018}, Gauss--Bonnet dark energy was treated as a holographic fluid interacting with the horizon. In contrast, the present study considers a Ricci--Gauss--Bonnet dark energy model and performs a scalar-field reconstruction within Ho\v{r}ava--Lifshitz cosmology, leading to a dynamical late-time behaviour. The paper is organized as follows. In Sec.~I, we present the motivation for studying dark energy within modified gravity frameworks and review the observational and theoretical background relevant to Ricci--Gauss--Bonnet dark energy and Hořava--Lifshitz cosmology. Sec.~II is devoted to a brief overview of Hořava--Lifshitz gravity, where the ADM formalism, the gravitational action, and the modified Friedmann equations are introduced, followed by the definitions of the effective dark-energy density and pressure in a spatially flat FLRW universe. In Sec.~III, we construct the Ricci--Gauss--Bonnet dark energy model and perform a detailed scalar-field reconstruction by deriving the explicit forms of the scalar-field kinetic term and the reconstructed potential. The dynamical evolution of the reconstructed scalar field is analyzed through the behavior of the potential gradient, the effective equation-of-state parameter, the parametric potential, and the phase-space trajectories. Sec.~IV is devoted to the physical viability of the model, where we examine classical stability via the squared speed of sound and test the validity of the generalized second law of thermodynamics at the apparent horizon. Finally, in Sec.~V, we summarize the main results and discuss the physical implications of the Ricci--Gauss--Bonnet dark energy model within the Hořava--Lifshitz cosmological framework.

\section{Ho\v{r}ava--Lifshitz Cosmology}

\subsection{ADM formalism and action}
The Arnowitt–Deser–Misner (ADM) formalism provides a natural framework for formulating modified gravity theories by explicitly separating space and time, thereby facilitating a Hamiltonian description of gravitational dynamics. Within this approach, modified gravity models have been constructed to explore deviations from General Relativity while preserving a well-defined constraint structure and dynamical consistency \cite{Gao2010}. Further developments have shown that the ADM representation is particularly suitable for the Hamiltonian analysis of covariant and renormalizable gravity theories, allowing a systematic investigation of their degrees of freedom and stability properties \cite{Chaichian2011}. These ideas form the conceptual backbone of Ho\v{r}ava--Lifshitz gravity, where anisotropic scaling between space and time is introduced at high energies to achieve power-counting renormalizability, leading to rich cosmological and gravitational phenomenology \cite{Horava2009}. Ho\v{r}ava--Lifshitz gravity has the metric written as \cite{Sotiriou2011}
\begin{equation}
ds^{2} = -N^{2} dt^{2} + g_{ij} (dx^{i} + N^{i} dt)(dx^{j} + N^{j} dt),
\end{equation}
where $N$ is the lapse function, $N^{i}$ is the shift vector, and $g_{ij}$ is the spatial metric.

The extrinsic curvature is defined as  \cite{Sotiriou2011}
\begin{equation}
K_{ij}
=
\frac{1}{2N}
\left(
\dot g_{ij}
-
\nabla_i N_j
-
\nabla_j N_i
\right),
\qquad
K=g^{ij}K_{ij}.
\end{equation}

The action of HL gravity is given by \cite{Sotiriou2011}
\begin{equation}
S
=
\int dt\, d^{3}x \, N \sqrt{g}
\left[
\frac{2}{\kappa^{2}}
\left(
K_{ij}K^{ij}
-
\lambda K^{2}
\right)
-
\mathcal{V}(g_{ij})
\right]
+
S_{\rm matter},
\end{equation}
where $\lambda$ is a dimensionless running coupling parameter and
$\mathcal{V}(g_{ij})$ is the potential term constructed from spatial curvature
invariants. The potential $\mathcal{V}(g_{ij})$ can be written in a generic form as a linear
combination of all curvature invariants constructed from the spatial metric,
consistent with foliation-preserving diffeomorphisms
\cite{Horava2009,Kiritsis2009,Mukohyama2010}. Decomposing the gravitational action into a kinetic and a potential part as \cite{Saridakis2010}
\begin{equation}
S_g=\int dt\, d^{3}x \, \sqrt{g}\, N \left( \mathcal{L}_{K}+\mathcal{L}_{V} \right),
\end{equation}
and under the assumption of detailed balance \cite{Horava2009}, which apart from
reducing the possible terms in the Lagrangian also allows for a quantum
inheritance principle \cite{Horava2009} (the $(D+1)$-dimensional theory acquires
the renormalization properties of the $D$-dimensional one), the full action of
Ho\v{r}ava--Lifshitz gravity is given by \cite{Saridakis2010}
\begin{align}
S_g
=
\int dt\, d^{3}x \, \sqrt{g}\, N
\Bigg[
&\frac{2}{\kappa^{2}}
\left(
K_{ij}K^{ij}
-
\lambda K^{2}
\right)
-
\frac{\kappa^{2}}{2 w^{4}} C_{ij}C^{ij}
+
\frac{\kappa^{2}\mu}{2 w^{2}}
\frac{\epsilon^{ijk}}{\sqrt{g}}
R_{i\ell} \nabla_{j} R^{\ell}{}_{k}
\nonumber\\
&\quad
-
\frac{\kappa^{2}\mu^{2}}{8} R_{ij}R^{ij}
+
\frac{\kappa^{2}\mu^{2}}{8(1-3\lambda)}
\left(
\frac{1-4\lambda}{4} R^{2}
+
\Lambda R
-
3\Lambda^{2}
\right)
\Bigg],
\label{actionHL}
\end{align}
where
\begin{equation}
K_{ij}
=
\frac{1}{2N}
\left(
\dot g_{ij}
-
\nabla_i N_j
-
\nabla_j N_i
\right),
\label{extrinsic}
\end{equation}
is the extrinsic curvature, and
\begin{equation}
C^{ij}
=
\frac{\epsilon^{ijk}}{\sqrt{g}}
\nabla_{k}
\left(
R_{i}^{j}
-
\frac{1}{4} R \delta_{i}^{j}
\right)
\label{cotton}
\end{equation}
is the Cotton tensor.

As in this study we are focusing on cosmological frameworks, we impose a Friedmann--Robertson--Walker (FRW) metric \cite{Saridakis2010},
\begin{equation}
N = 1, \qquad g_{ij} = a^{2}(t)\,\gamma_{ij}, \qquad N^{i}=0,
\label{FRWmetric}
\end{equation}
with
\begin{equation}
\gamma_{ij} dx^{i} dx^{j}
=
\frac{dr^{2}}{1-k r^{2}}
+
r^{2} d\Omega^{2},
\label{spatialmetric}
\end{equation}
where $k=-1,\,0,\,1$ correspond to an open, flat, and closed universe, respectively. In the above expressions, all covariant derivatives are defined with respect to the spatial metric $g_{ij}$. The tensor $\epsilon^{ijk}$ denotes the totally antisymmetric unit tensor, $\lambda$ is a dimensionless constant, and $\Lambda$ is a negative constant which is related to the cosmological constant in the infrared (IR) limit \cite{Saridakis2010}.

\subsection{Modified Friedmann equations}
The equations of motion in this scenario are the following \cite{Saridakis2010}
\begin{align}
H^{2}
&=
\frac{\kappa^{2}}{6(3\lambda-1)}
\left[
\frac{3\lambda-1}{4}\,\dot{\phi}^{2}
+
V_{\phi}(\phi)
\right]
+
\frac{\kappa^{2}}{6(3\lambda-1)}
\left[
\frac{3\lambda-1}{4}\,\dot{\sigma}^{2}
+
V_{\sigma}(\sigma)
\right]
\nonumber\\
&\quad
-
\frac{3\kappa^{2}\mu^{2}k^{2}}{8(3\lambda-1)a^{4}}
-
\frac{3\kappa^{2}\mu^{2}\Lambda^{2}}{8(3\lambda-1)}
+
\frac{\kappa^{4}\mu^{2}\Lambda k}{8(3\lambda-1)^{2}a^{2}},
\label{HLfriedmann1}
\\[2mm]
\dot{H}
+
\frac{3}{2}H^{2}
&=
-
\frac{\kappa^{2}}{4(3\lambda-1)}
\left[
\frac{3\lambda-1}{4}\,\dot{\phi}^{2}
-
V_{\phi}(\phi)
\right]
-
\frac{\kappa^{2}}{4(3\lambda-1)}
\left[
\frac{3\lambda-1}{4}\,\dot{\sigma}^{2}
-
V_{\sigma}(\sigma)
\right]
\nonumber\\
&\quad
-
\frac{\kappa^{2}\mu^{2}k^{2}}{8(3\lambda-1)a^{4}}
+
\frac{3\kappa^{2}\mu^{2}\Lambda^{2}}{8(3\lambda-1)}
+
\frac{\kappa^{4}\mu^{2}\Lambda k}{16(3\lambda-1)^{2}a^{2}} .
\label{HLfriedmann2}
\end{align}

Here, $V_{\sigma}(\sigma)$ denotes the potential term associated with the
$\sigma$-field. The function $V_{\phi}(\phi)$ acts as the potential term for the $\phi$-field. 

\subsection{Effective dark-energy density and pressure}
Following \cite{Saridakis2010}, the effective dark-energy density and pressure in Ho\v{r}ava--Lifshitz cosmology are defined as Eqs.~(\ref{rhoDE}) and (\ref{pDE})
\begin{align}
\rho_{\rm DE}
&=
\frac{3\lambda-1}{4}\,\dot{\sigma}^{2}
+
V_{\sigma}(\sigma)
-
\frac{3\kappa^{2}\mu^{2}k^{2}}{8(3\lambda-1)a^{4}}
-
\frac{3\kappa^{2}\mu^{2}\Lambda^{2}}{8(3\lambda-1)},
\label{rhoDE}
\\[2mm]
p_{\rm DE}
&=
\frac{3\lambda-1}{4}\,\dot{\sigma}^{2}
-
V_{\sigma}(\sigma)
-
\frac{\kappa^{2}\mu^{2}k^{2}}{8(3\lambda-1)a^{4}}
+
\frac{3\kappa^{2}\mu^{2}\Lambda^{2}}{8(3\lambda-1)} .
\label{pDE}
\end{align}

For a spatially flat universe ($k=0$), the dark-energy density \eqref{rhoDE} and pressure \eqref{pDE} take the form
reduce to
\begin{align}
\rho_{\rm DE}
&=
\frac{3\lambda-1}{4}\,\dot{\sigma}^{2}
+
V_{\sigma}(\sigma)
-
\frac{3\kappa^{2}\mu^{2}\Lambda^{2}}{8(3\lambda-1)},
\label{rhoDEk0}
\\[2mm]
p_{\rm DE}
&=
\frac{3\lambda-1}{4}\,\dot{\sigma}^{2}
-
V_{\sigma}(\sigma)
+
\frac{3\kappa^{2}\mu^{2}\Lambda^{2}}{8(3\lambda-1)} .
\label{pDEk0}
\end{align}
From Equations \eqref{rhoDEk0} and \eqref{pDEk0}
\begin{equation}
\rho_{\rm DE}+p_{\rm DE}
=
\frac{3\lambda-1}{2}\,\dot{\sigma}^{2}.
\label{addition1}
\end{equation}

\begin{equation}
\rho_{\rm DE}-p_{\rm DE}
=
2V_{\sigma}(\sigma)
-
\frac{3\kappa^{2}\mu^{2}\Lambda^{2}}{4(3\lambda-1)}.
\label{subtraction1}
\end{equation}

The above formulation establishes the cosmological field equations of
Ho\v{r}ava--Lifshitz gravity in a spatially flat FLRW background and provides the effective dark-energy density and pressure in terms of the scalar-field sector. This framework provides a consistent setting in which modified gravity effects can be interpreted as an effective dark energy fluid driving cosmic dynamics.

Having set up the general cosmological equations in Ho\v{r}ava--Lifshitz
gravity, we are now in a position to specify the form of the dark-energy
density. In the next section, we are going to introduce a Ricci--Gauss--Bonnet dark energy model and we would present a reconstruct scheme for an effective scalar-field description within the Ho\v{r}ava--Lifshitz cosmological framework along with exploring its dynamical and thermodynamical implications in detail.

\section{Ricci--Gauss--Bonnet Dark Energy}
\textcolor{black}{In contrast to earlier studies of Ricci--Gauss--Bonnet dark energy in standard FLRW cosmology, the present work incorporates the RGB sector within the Ho\v{r}ava--Lifshitz framework. In this setting, the modified gravitational dynamics alter the reconstruction procedure and influence the resulting stability and thermodynamic behaviour. Therefore, the present formulation provides a distinct realization of curvature-based dark energy beyond conventional Einstein gravity.}

\subsection{Ricci–Gauss–Bonnet Dark Energy Density}
This section is devoted to the modeling of Ricci--Gauss--Bonnet (RGB) dark energy, a concept originally proposed in \cite{Saridakis2018} and further utilized in \cite{Iqbal2018, Dubey2025}, within the framework of Ho\v{r}ava--Lifshitz cosmology. This is motivated by holographic considerations and higher-curvature corrections arising in modified gravity theories. Based on this consideration, the RGB dark energy density is constructed as a linear combination of the Ricci scalar and the Gauss-Bonnet invariant. Employing a spatially flat FLRW background and assuming a power-law form of scale factor, we derive explicit expressions for the effective dark energy density and pressure. This approach helps us to formulate a consistent reconstruction of the RGB dark energy sector as an effective scalar-field description in Ho\v{r}ava--Lifshitz cosmology, providing a convenient framework to analyze the resulting cosmic evolution and dynamical behavior. We define the Ricci--Gauss--Bonnet dark energy density as
\begin{equation}
\rho_{DE} = 3 c^{2}\left(\alpha R + \beta \mathcal{G}\right),
\end{equation}
where $c^{2}$ is a dimensionless parameter, $\alpha$ and $\beta$ are constants, $R$ is the Ricci scalar, and $\mathcal{G}$ is the Gauss--Bonnet invariant. \textcolor{black}{Let us note that the parameters appearing in the above equation have physical interpretations within the Ricci–Gauss–Bonnet dark energy framework. Firstly, $c$, which is a dimensionless constant, plays a role that is analogous to the parameter appearing in holographic dark energy models and determines the overall strength of the effective dark-energy density associated with the curvature sector. The influence of local curvature effects that are typically associated with the late-time cosmic expansion is characterized by $\alpha$. Furthermore, the parameter $\beta$ controls the contribution coming from the higher-curvature Gauss–Bonnet corrections.  Such corrections are generally expected to be more significant at the earlier stages. Therefore, the combined effect of the parameters $\alpha$ and $\beta$ determines the relative importance of the Ricci-driven and Gauss–Bonnet-driven dynamics in the reconstructed dark-energy sector.}

For the FLRW background, these quantities read
\begin{equation}
R = 6(2H^{2} + \dot{H}),
\end{equation}
\begin{equation}
\mathcal{G} = 24H^{2}(H^{2} + \dot{H}).
\end{equation}

Substituting these expressions into the dark energy density, we obtain
\begin{equation}
\rho_{DE} = 18c^{2}\alpha(2H^{2} + \dot{H}) + 72c^{2}\beta H^{2}(H^{2} + \dot{H}).
\label{rhoRGBexplicit}
\end{equation}

At this juncture, it may be noted that the Ricci--Gauss--Bonnet holographic dark energy model proposed in
\cite{Saridakis2018}, where it is defined as the dark energy density through a holographic description in which the infrared cutoff is determined directly by curvature invariants, $L^{-2}=-\alpha R+\beta\sqrt{|\mathcal{G}|}$ \cite{Saridakis2018} within the standard FRW cosmology. On the other hand, the present study considers the Ricci--Gauss--Bonnet combination as an effective curvature-induced dark energy embedded in Ho\v{r}ava--Lifshitz gravity, where higher-curvature terms arise naturally from the modified gravitational framework and are reconstructed into an effective scalar-field description. The Gauss--Bonnet contribution is dominant in the early phase of the universe and the Ricci term influences the late-time accelerated expansion. Hence, the underlying physical interpretation differs fundamentally from the holographic construction. The two approaches i.e. one presented in \cite{Saridakis2018} and the current work, therefore, share a common geometric motivation, but they represent distinct realizations of Ricci--Gauss--Bonnet dark energy. The form adopted here has earlier been utilized in \cite{Altaibayeva2025}.

To determine the corresponding dark-energy pressure, we employ the continuity
equation for the effective dark-energy fluid,
\begin{equation}
\dot{\rho}_{\rm DE}
+
3H\left(\rho_{\rm DE}+p_{\rm DE}\right)=0.
\label{continuityRGB}
\end{equation}
Solving Eq.~\eqref{continuityRGB} for the pressure yields
\begin{equation}
p_{\rm DE}
=
-\rho_{\rm DE}
-
\frac{1}{3H}\dot{\rho}_{\rm DE}.
\label{pfromrho}
\end{equation}

Assuming a power-law scale factor
\begin{equation}
a(t)=a_{0}t^{n}, \qquad n>1,
\label{powerlaw}
\end{equation}
the Hubble parameter and its derivative become
\begin{equation}
H=\frac{n}{t},
\qquad
\dot{H}=-\frac{n}{t^{2}}.
\end{equation}

\textcolor{black}{At this juncture, it should be noted that the power-law scale factor $a(t)=a_0 t^n$ is adopted.  It is a commonly used phenomenological ansatz in cosmological studies. Such forms are frequently employed in the literature to obtain analytical insight into the behaviour of dark-energy models and modified gravity scenarios. In the present work this assumption is used to explore the qualitative features of the reconstructed Ricci–Gauss–Bonnet dark-energy sector within the Ho\v{r}ava--Lifshitz cosmological framework. Therefore, the background expansion is not intended to represent a unique dynamical solution derived from the modified Friedmann equations. Instead, it serves as a convenient illustrative choice that allows us to analyse the scalar-field reconstruction and its physical implications in the HL context.}

Substituting these expressions into Eq.~\eqref{rhoRGBexplicit}, the dark-energy
density can be written explicitly as a function of cosmic time:
\begin{equation}
\rho_{\rm DE}
=
\frac{18c^{2}\alpha(2n^{2}-n)}{t^{2}}
+
\frac{72c^{2}\beta n^{2}(n^{2}-n)}{t^{4}}.
\label{rhoRGBt}
\end{equation}
Taking the time derivative of Eq.~\eqref{rhoRGBt}, we find
\begin{equation}
\dot{\rho}_{\rm DE}
=
-\frac{36c^{2}\alpha(2n^{2}-n)}{t^{3}}
-
\frac{288c^{2}\beta n^{2}(n^{2}-n)}{t^{5}}.
\label{rhodotRGB}
\end{equation}

Finally, substituting Eqs.~\eqref{rhoRGBt} and \eqref{rhodotRGB} into
Eq.~\eqref{pfromrho}, and using $H=n/t$, the dark-energy pressure is obtained as
\begin{align}
p_{\rm DE}
&=
\frac{6c^{2}\alpha(2n^{2}-n)}{t^{2}}
\left(
\frac{2}{n}-3
\right)
+
\frac{24c^{2}\beta n^2(n-1)(4-3n)}{t^{4}} .
\label{pRGBsimplified}
\end{align}

Equations~\eqref{rhoRGBt} and \eqref{pRGBsimplified} fully characterize the effective
dark-energy fluid associated with the Ricci--Gauss--Bonnet model under the
assumption of a power-law cosmic expansion. These expressions can be used to
study the evolution of the equation-of-state parameter
$\omega_{\rm DE}=p_{\rm DE}/\rho_{\rm DE}$ and the resulting cosmological
dynamics. From Equations~\eqref{rhoRGBt} and \eqref{pRGBsimplified} we have

\begin{equation}
\rho_{\rm DE}+p_{\rm DE}
=
\frac{12c^{2}\alpha(2n-1)}{t^{2}}
+
\frac{24c^{2}\beta n^{2}(n-1)}{t^{4}} .
\label{addition2}
\end{equation}

\begin{equation}
\rho_{\rm DE}-p_{\rm DE}
=
\frac{12c^{2}\alpha(2n^{2}-n)}{t^{2}}
\left(3-\frac{1}{n}\right)
+
\frac{48c^{2}\beta n^{2}(n-1)(3n-2)}{t^{4}} .
\label{subtraction2}
\end{equation}

Equating \eqref{addition1} and \eqref{addition2} we get

\begin{equation}
\dot{\sigma}^{2}
=
\frac{24c^{2}\alpha(2n-1)}{(3\lambda-1)\,t^{2}}
+
\frac{48c^{2}\beta n^{2}(n-1)}{(3\lambda-1)\,t^{4}} .
\label{31}
\end{equation}

Equating \eqref{subtraction1} and \eqref{subtraction2} we get 

\begin{equation}
V_{\sigma}(\sigma(t))
=
\frac{6c^{2}\alpha(2n^{2}-n)}{t^{2}}
\left(3-\frac{1}{n}\right)
+
\frac{24c^{2}\beta n^{2}(n-1)(3n-2)}{t^{4}}
+
\frac{3\kappa^{2}\mu^{2}\Lambda^{2}}{8(3\lambda-1)} .
\label{32}
\end{equation}

Equations \eqref{31} and \eqref{32} demonstrate that the Ricci--Gauss--Bonnet dark energy
sector can be consistently reconstructed as an effective scalar-field model
within Ho\v{r}ava--Lifshitz cosmology, with higher-curvature contributions
governing early-time dynamics and Ricci-type terms driving late-time cosmic
acceleration. If we begin from Eq.~(31), we define the constants
\begin{equation}
A \equiv \frac{24c^{2}\alpha(2n-1)}{3\lambda-1},
\qquad
B \equiv \frac{48c^{2}\beta n^{2}(n-1)}{3\lambda-1}.
\end{equation}
Equation~\eqref{31} can then be written as
\begin{equation}
\dot{\sigma}^{2}
=
\frac{A}{t^{2}}+\frac{B}{t^{4}}
=
\frac{A t^{2}+B}{t^{4}},
\end{equation}
which yields
\begin{equation}
\dot{\sigma}
=
\frac{\sqrt{A t^{2}+B}}{t^{2}}.
\label{sigmadot}
\end{equation}

Integrating Eq.~\eqref{sigmadot} with respect to cosmic time,
\begin{equation}
\sigma(t)
=
\int \frac{\sqrt{A t^{2}+B}}{t^{2}}\,dt .
\label{sigmaintegral}
\end{equation}

To evaluate the integral, we rewrite the integrand as
\begin{equation}
\frac{\sqrt{A t^{2}+B}}{t^{2}}
=
\frac{1}{t}\sqrt{A+\frac{B}{t^{2}}}.
\end{equation}
Introducing the substitution
\begin{equation}
u=\sqrt{A t^{2}+B},
\qquad
du=\frac{A t}{\sqrt{A t^{2}+B}}\,dt,
\end{equation}
the integral in Eq.~\eqref{sigmaintegral} can be computed straightforwardly.
After integration, one obtains
\begin{equation}
\sigma(t)
=
\frac{1}{2}
\left[
-\frac{\sqrt{B + A t^{2}}}{t^{2}}
+
\frac{A}{\sqrt{B}}
\ln\!\left(
\frac{t}{\sqrt{B}\left(\sqrt{B}+\sqrt{B+A t^{2}}\right)}
\right)
\right]
+
\sigma_{0},
\label{sigmasolution}
\end{equation}
where \(\sigma_{0}\) is an integration constant.

\subsection{The reconstruction scheme}

Using the scalar-field equation of motion obtained from the continuity equation,
\begin{equation}
\ddot{\sigma}
+
\frac{3n}{t}\dot{\sigma}
+
\frac{2}{3\lambda-1}\frac{dV_{\sigma}}{d\sigma}
=0 ,
\label{eq:eom_sigma}
\end{equation}
one can express the derivative of the potential as
\begin{equation}
\frac{dV_{\sigma}}{d\sigma}
=
-\frac{3\lambda-1}{2}
\left(
\ddot{\sigma}
+
\frac{3n}{t}\dot{\sigma}
\right).
\label{eq:dVdsigma_general}
\end{equation}

From Eq.~\eqref{sigmadot}, the first derivative of the scalar field is given by
\begin{equation}
\dot{\sigma}
=
\frac{1}{t}\sqrt{A+\frac{B}{t^{2}}}
=
\frac{\sqrt{A t^{2}+B}}{t^{2}},
\label{eq:sigmadot}
\end{equation}
which upon differentiation yields
\begin{equation}
\ddot{\sigma}
=
\frac{A}{t\sqrt{A t^{2}+B}}
-
\frac{2\sqrt{A t^{2}+B}}{t^{3}} .
\label{eq:sigmaddot}
\end{equation}

Substituting Eqs.~(\ref{eq:sigmadot}) and (\ref{eq:sigmaddot}) into
Eq.~(\ref{eq:dVdsigma_general}), we obtain
\begin{equation}
\frac{dV_{\sigma}}{d\sigma}
=
-\frac{3\lambda-1}{2}
\left[
\frac{A}{t\sqrt{A t^{2}+B}}
+
\frac{(3n-2)\sqrt{A t^{2}+B}}{t^{3}}
\right].
\label{eq:dVdsigma_final}
\end{equation}

Equation~(\ref{eq:dVdsigma_final}) provides the explicit time-dependent form
of the gradient of the reconstructed scalar-field potential corresponding to
the Ricci--Gauss--Bonnet dark energy sector in Ho\v{r}ava--Lifshitz cosmology. As already mentioned, in \eqref{eq:dVdsigma_final}, we have $A \equiv \frac{24c^{2}\alpha(2n-1)}{3\lambda-1},
\qquad
B \equiv \frac{48c^{2}\beta n^{2}(n-1)}{3\lambda-1}$.

\begin{figure}[t]
    \centering

    \begin{subfigure}[t]{0.48\textwidth}
        \centering
        \includegraphics[width=\linewidth]{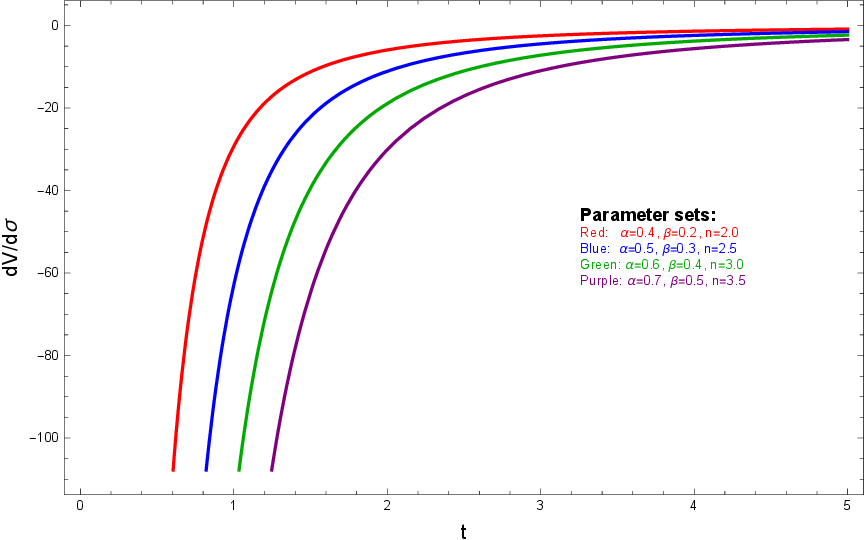}
        \caption{
        Evolution of $\mathrm{d}V_{\sigma}/\mathrm{d}\sigma$ as a function of cosmic time $t$
        for different parameter choices.
        }
        \label{fig:dVdsigma}
    \end{subfigure}
    \hfill
    \begin{subfigure}[t]{0.48\textwidth}
        \centering
        \includegraphics[width=\linewidth]{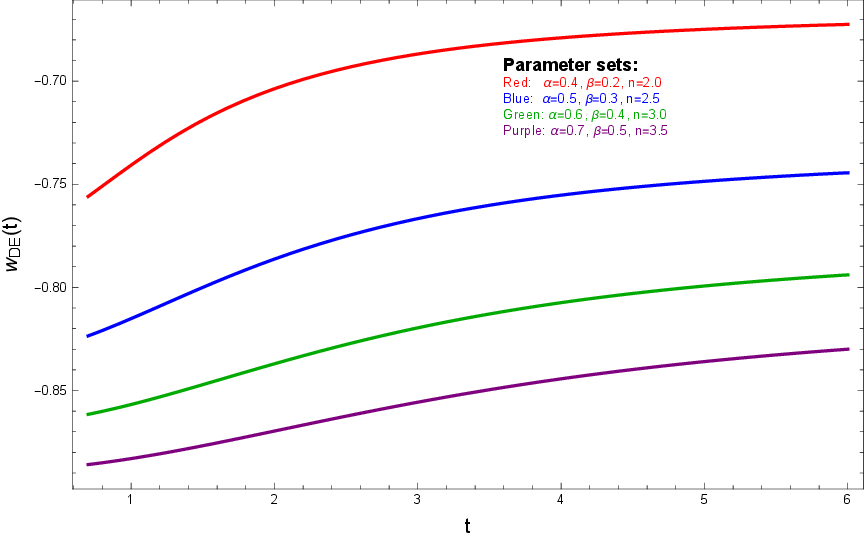}
        \caption{
        Evolution of the effective dark-energy equation-of-state parameter
        $\omega_{\rm DE}(t)$ for different model parameters.
        }
        \label{fig:wDE}
    \end{subfigure}

    \vspace{0.35cm}

    \begin{subfigure}[t]{0.48\textwidth}
        \centering
        \includegraphics[width=\linewidth]{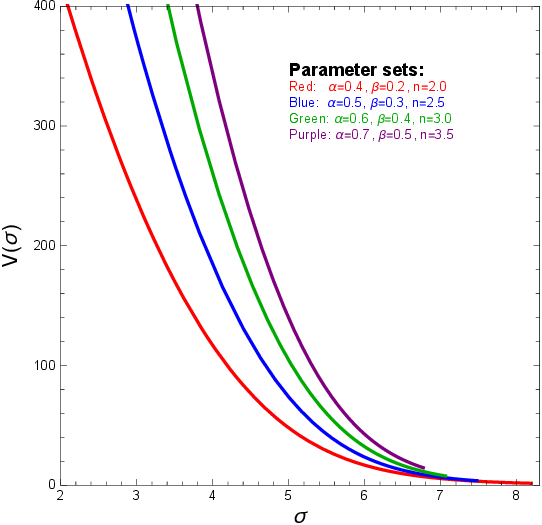}
        \caption{
        Parametric plot of the reconstructed scalar-field potential $V(\sigma)$
        obtained by eliminating the cosmic time $t$.
        }
        \label{fig:Vsigma}
    \end{subfigure}
    \hfill
    \begin{subfigure}[t]{0.48\textwidth}
        \centering
        \includegraphics[width=\linewidth]{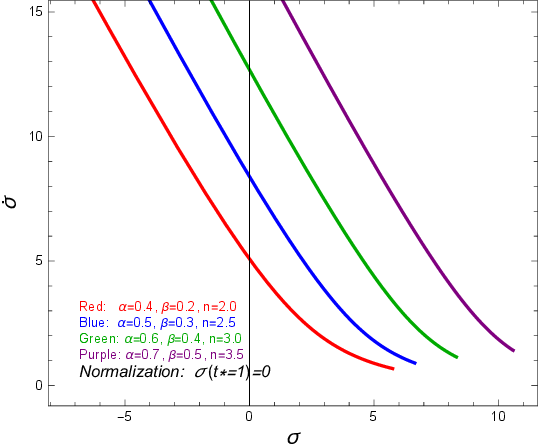}
        \caption{
        Phase-space trajectories in the $(\sigma,\dot{\sigma})$ plane
        for the reconstructed scalar field.
        }
        \label{fig:phasespace}
    \end{subfigure}

    \caption{
    Reconstructed scalar-field dynamics in the Ricci--Gauss--Bonnet dark energy
    model within Ho\v{r}ava--Lifshitz cosmology.
    Panels show:
    (a) $\mathrm{d}V_{\sigma}/\mathrm{d}\sigma$ versus cosmic time,
    (b) evolution of the effective equation-of-state parameter $\omega_{\rm DE}$,
    (c) parametric reconstruction of the potential $V(\sigma)$,
    and (d) phase-space trajectories in the $(\sigma,\dot{\sigma})$ plane.
    In all cases, $c=1$ and $\lambda=1.2$ are fixed, while the parameter sets
    $(\alpha,\beta,n)$ correspond to
    $(0.4,0.2,2.0)$ (red),
    $(0.5,0.3,2.5)$ (blue),
    $(0.6,0.4,3.0)$ (green),
    and $(0.7,0.5,3.5)$ (purple).
    }
    \label{fig:combined_panels}
\end{figure}

Figure~\ref{fig:dVdsigma} depicts the evolution of the derivative of the reconstructed scalar-field potential, $\mathrm{d}V_{\sigma}/\mathrm{d}\sigma$, as a function of cosmic
time $t$ for different choices of the model parameters $(\alpha,\beta,n)$, obtained from Eq.~(\ref{eq:dVdsigma_final}). For all parameter sets, the potential gradient is negative at early times. With evolution of the universe, it exhibits monotone increasing pattern approaching towards zero. This indicates a smooth rolling behavior of the effective scalar field without any intrinsic divergence for $t>0$. 
A further insight into the plots reveals that at early times, the magnitude of $\mathrm{d}V_{\sigma}/\mathrm{d}\sigma$ is larger. This can be interpreted as the dominance of the Gauss--Bonnet contribution proportional to the parameter $\beta$.  Here, $\beta$ is thought of as improving the higher-curvature effects in the ultraviolet regime. As cosmic time increases, the curves gradually flatten and approach zero. This gradual flattening is indicating a weakening of the scalar-field driving force and a transition toward Ricci-dominated dynamics, which are primarily governed by the parameter $\alpha$. In the parameter set, we increase the power-law index $n$ shifts the curves downward and delays the approach to the asymptotic regime, demonstrating the sensitivity of the reconstructed potential
gradient to the background expansion rate. Considering all the above, Fig.~\ref{fig:dVdsigma} leads us to conclude that the Ricci--Gauss--Bonnet dark energy sector can be consistently reconstructed as an effective scalar-field description within Ho\v{r}ava--Lifshitz cosmology. The smooth and finite behavior of
$\mathrm{d}V_{\sigma}/\mathrm{d}\sigma$ for all the sets of parameters considered here indicate the dynamical viability of the model and is indicative of the role of higher-curvature corrections associated with the scalar-field dynamics across different cosmic epochs.

Using Eqs.~\eqref{rhoDEk0} and \eqref{pDEk0}, the effective dark-energy equation-of-state parameter
is defined as
\begin{equation}
\omega_{\rm DE}
=
\frac{p_{\rm DE}}{\rho_{\rm DE}}
=
\frac{\dfrac{3\lambda-1}{4}\dot{\sigma}^{2}
-
V_{\sigma}(\sigma)
+
\dfrac{3\kappa^{2}\mu^{2}\Lambda^{2}}{8(3\lambda-1)}
}{
\dfrac{3\lambda-1}{4}\dot{\sigma}^{2}
+
V_{\sigma}(\sigma)
-
\dfrac{3\kappa^{2}\mu^{2}\Lambda^{2}}{8(3\lambda-1)}
}.
\label{eq:wDE_def}
\end{equation}

Substituting $\dot{\sigma}^{2}$ from Eq.~(31) and the reconstructed scalar
potential $V_{\sigma}(\sigma(t))$ from Eq.~(32), we obtain
\begin{equation}
\omega_{\rm DE}(t)
=
\frac{
\displaystyle
\frac{3\lambda-1}{4}
\left(
\frac{A}{t^{2}}+\frac{B}{t^{4}}
\right)
-
\left[
6c^{2}\alpha(2n^{2}-n)\frac{1}{t^{2}}
\left(3-\frac{1}{n}\right)
+
24c^{2}\beta n^{2}(n-1)(3n-2)\frac{1}{t^{4}}
+
\frac{3\kappa^{2}\mu^{2}\Lambda^{2}}{8(3\lambda-1)}
\right]
+
\frac{3\kappa^{2}\mu^{2}\Lambda^{2}}{8(3\lambda-1)}
}{
\displaystyle
\frac{3\lambda-1}{4}
\left(
\frac{A}{t^{2}}+\frac{B}{t^{4}}
\right)
+
\left[
6c^{2}\alpha(2n^{2}-n)\frac{1}{t^{2}}
\left(3-\frac{1}{n}\right)
+
24c^{2}\beta n^{2}(n-1)(3n-2)\frac{1}{t^{4}}
+
\frac{3\kappa^{2}\mu^{2}\Lambda^{2}}{8(3\lambda-1)}
\right]
-
\frac{3\kappa^{2}\mu^{2}\Lambda^{2}}{8(3\lambda-1)}
}.
\label{eq:wDE_full}
\end{equation}

Here the constants $A$ and $B$ are defined as
\begin{equation}
A \equiv \frac{24c^{2}\alpha(2n-1)}{3\lambda-1},
\qquad
B \equiv \frac{48c^{2}\beta n^{2}(n-1)}{3\lambda-1}.
\label{eq:ABdef}
\end{equation}

The effective equation-of-state parameter $\omega_{\rm DE}$ is obtained by substituting the reconstructed scalar-field kinetic term $\dot{\sigma}^{2}$ and the potential $V_{\sigma}(\sigma(t))$ into the effective dark-energy density and pressure, resulting in the explicit time-dependent expression given in Eq.~\eqref{eq:wDE_full} and is plotted in Fig.~\ref{fig:wDE}. As shown in Fig.~\ref{fig:wDE}, the equation-of-state parameter evolves smoothly with cosmic time and stays above the phantom boundary and for some choice of parameters it lies in the vicinity of the cosmological-constant limit. Figure~\ref{fig:wDE} further shows that $\omega_{\rm DE}$ does not asymptotically approach the
cosmological-constant value $\omega_{\rm DE}=-1$ at late times.  Instead, it deviates further away from it as $t$ increases.
This behavior can be interpreted to arise due to the explicit time dependence of both the kinetic term $\dot{\sigma}^{2}$ and the reconstructed potential $V_{\sigma}(\sigma)$
in Eq.~\eqref{eq:wDE_full}, which prevents the reach of a pure vacuum-dominated regime. 

In general we can say that at larger $t$ the residual contribution of the kinetic term, together with the Ricci-driven part of the potential, leads to a gradual departure of $\omega_{\rm DE}$ from the cosmological-constant value. From this we can say that a  dynamical dark-energy behavior is apparent from the Ho\v{r}ava--Lifshitz cosmology in Ricci--Gauss--Bonnet framework.

In Fig.~\ref{fig:Vsigma}, the smaller values of $\sigma$ correspond to the early epoch of the universe. During this phase, which corresponds to smaller values of $\sigma$, the reconstructed potential $V(\sigma)$ appears to be steeper.  The behaviour indicates the dominance of the Gauss--Bonnet contribution, which enters through higher-curvature terms proportional to $t^{-4}$. As $t$ increases, the effect of the Gauss--Bonnet sector gradually weakens. On the other hand, the Ricci curvature contribution becomes more important. This change is manifested as a smooth variation in the slope of $V(\sigma)$, indicating a continuous transition between the Gauss--Bonnet–dominated and
Ricci-dominated regimes. In the phase-space diagram presented in Fig. \ref{fig:phasespace} the smooth and non-closed trajectories indicate a monotonic evolution of
the scalar field without oscillatory behavior, confirming the regularity of the reconstructed dynamics. Different curves correspond to different choices of the model parameters $(\alpha,\beta,n)$.

\textcolor{black}{It may be noted that the reconstructed scalar field in the present work should be regarded as an effective description of the dark-energy sector rather than a fundamental degree of freedom arising directly from Ho\v{r}ava--Lifshitz theory. The reconstruction provides a convenient parametrization of the modified cosmological dynamics in terms of a scalar-field framework.}

\section{Stability and Thermodynamics}

The squared speed of sound, defined as $v_s^2 \equiv dp_{\rm DE}/d\rho_{\rm DE}$, is evaluated through the time derivatives of the effective
dark-energy pressure and density. Since both $p_{\rm DE}$ and $\rho_{\rm DE}$ depend explicitly on the cosmic time $t$ in the
Ho\v{r}ava--Lifshitz cosmological framework, the squared sound speed can be written as
\begin{equation}
v_s^{2}(t)=\frac{\dot p_{\rm DE}}{\dot\rho_{\rm DE}}
=
\frac{
6\alpha(2n^{2}-n)\!\left(\frac{2}{n}-3\right)t^{2}
+24\beta n^{2}(n-1)(4-3n)
}{
18\alpha(2n^{2}-n)t^{2}
+72\beta n^{2}(n^{2}-n)
}.
\label{eq:vs2}
\end{equation}
A positive value of $v_s^{2}$ indicates classical stability \cite{Jawad2019,RaniAshrafJawad2022,SergijenkoNovosyadlyj2015} of the effective Ricci--Gauss--Bonnet dark energy sector against small perturbations within Ho\v{r}ava--Lifshitz cosmology. In this framework, higher-curvature contributions inherent to Ho\v{r}ava--Lifshitz gravity modify the background dynamics, thereby influencing the propagation of dark-energy perturbations. At early times, the Gauss--Bonnet term dominates the behavior of $v_s^{2}$, reflecting the ultraviolet character of Ho\v{r}ava--Lifshitz gravity, whereas at late times the Ricci curvature contribution governs the asymptotic evolution, corresponding to the infrared limit of the theory.

\textcolor{black}{It is pertinent to note that a non-trivial constraint is imposed on the power-law index $n$ as well as on the coupling parameters by the condition $v_s^2 > 0$ imposes. To be specific, for an accelerating universe $(n > 1)$, the negative contribution arising from the first term in the numerator must be suitably counterbalanced by the Gauss--Bonnet sector. This indicates that the classical stability of the model is inherently regime-dependent, favouring a restricted window corresponding to the late-time evolution of the Universe.}

\begin{figure}[t]
    \centering
    \includegraphics[width=0.65\textwidth]{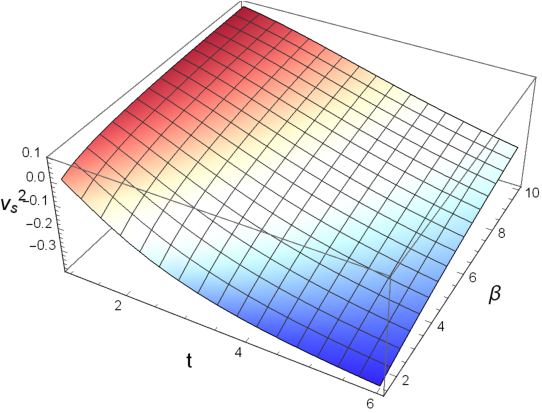}
    \caption{
    Three-dimensional evolution of the squared speed of sound $v_s^2$
    as a function of cosmic time $t$ and the Gauss--Bonnet coupling
    parameter $\beta$, for a representative accelerated expansion with
    $1<n<4/3$. The parameter $\alpha$ is fixed to a positive value to
    ensure a well-defined dark-energy sector. The positivity of $v_s^2$
    over the displayed parameter range indicates classical stability of
    the Ricci--Gauss--Bonnet dark energy model within
    Ho\v{r}ava--Lifshitz cosmology.
    }
    \label{fig:vs2_3D}
\end{figure}

Let us now have an insight into Eq. \eqref{eq:vs2} to inspect the conditions for stability. For an accelerating universe ($n>1$) with positive curvature couplings ($\alpha>0$, $\beta>0$), the denominator of $v_s^{2}$ necessarily remains positive
for all $t>0$. Thus, the stability condition $v_s^{2}>0$ depends entirely on the sign of the numerator. In order the first term of the numerator to be positive, we require $n<\frac{2}{3}$, which contradicts $n>1$. Thus, the first term is necessarily negative. Hence, in order to have $v_s^2>0$, the numerator positive, we need the second term of the numerator to be positive and dominate over the first term. For this, first of all, we need $1<n<\frac{4}{3}$ and $\beta$ needs to be sufficiently large to dominate over the first term. Hence, we can sum up by saying that the classical stability requires that the Gauss--Bonnet term dominates over the Ricci term, which is achieved for sufficiently large $\beta$ and moderate values of the power-law index $n$. In Fig. \ref{fig:vs2_3D}, we have displayed the evolution of the squared speed of sound $v_s^2$ as a function of cosmic time $t$ and the Gauss--Bonnet coupling parameter $\beta$. In alignment with our discussion, we observed that for lower values of $\beta$, the $v_s^2$ is negative, but gradually transitions to positive as the $\beta$ values increase. Moreover, we observe that the surface is smooth and there is a clear monotone pattern and there is no singularity or divergence. 

\textcolor{black}{A positive value of $v_s^2$ indicates classical stability of the effective Ricci--Gauss--Bonnet dark energy sector against small perturbations within Ho\v{r}ava--Lifshitz cosmology. In this framework, the higher-curvature contributions inherent to Ho\v{r}ava--Lifshitz gravity modify the background dynamics and thereby influence the propagation of perturbations in the dark-energy sector. At early times, the Gauss--Bonnet contribution plays a dominant role, reflecting the ultraviolet behaviour of the theory, whereas at late times the Ricci curvature term governs the asymptotic evolution corresponding to the infrared regime.}

\textcolor{black}{However, it may be noted that the above criterion provides only a necessary condition for classical stability. It may be mentioned that a more comprehensive analysis would require a detailed investigation of scalar perturbations within the Ho\v{r}ava--Lifshitz framework, including the possible occurrence of ghost instabilities as well as gradient instabilities in the scalar sector. Such effects may play an important role in modified gravity theories where additional degrees of freedom are present. A full perturbative investigation, however, lies beyond the scope of the present work and is therefore left for future study.}

\subsection{Test for generalized second law of thermodynamics}
For cosmological models, the generalized second law of thermodynamics (GSLT) provides us with a fundamental criterion for consistency \cite{Nojiri2004,Nojiri2020,Odintsov2024,Odintsov2025}. The GSLT requires the total entropy of the universe to be a non-decreasing function of cosmic time $t$. The total entropy consists of the horizon entropy and the entropy of the cosmic fluid enclosed by it \cite{Shekh2020}.  Since the total entropy is a non-decreasing function of cosmic time, the time derivative of the total entropy has to be non-negative in order for GSLT to hold \cite{Shekh2020}. The thermodynamical consistency of emergent cosmological dynamics based on generalized entropies has been examined in detail \cite{Luciano2023}. In modified gravity and dynamical dark energy scenarios, the validity of the GSLT is not guaranteed a priori and must be examined explicitly \cite{Bamba2011,Sharif2012,Sharif2013,Tian2015,Bamba2016,Myrzakulov2013}. We investigate the validity of the generalized second law of thermodynamics by considering the entropy evolution of the apparent horizon and the cosmic fluid. Our results indicate that the generalized second law holds for a wide range of model parameters. In order to examine the validity of the generalized second law of thermodynamics (GSLT) in the Ricci--Gauss--Bonnet (RGB) dark energy model within the framework of Ho\v{r}ava--Lifshitz cosmology, we consider the apparent horizon as the thermodynamically relevant boundary of the universe. We consider a spatially flat FLRW universe with a power-law scale factor $a(t)=a_0 t^{n},~ n>1$, which leads to an accelerated expansion. The Hubble parameter and apparent horizon radius are given by
\begin{equation}
H=\frac{n}{t}, \qquad R_A=\frac{1}{H}=\frac{t}{n}.
\end{equation}

\textcolor{black}{We would like to mention that the thermodynamical properties of the cosmological horizon can be modified in theories of gravity beyond General Relativity. In Hořava–Lifshitz gravity the gravitational action contains higher–order spatial curvature terms and anisotropic scaling between space and time, which modify the background cosmological dynamics. These modifications can affect the thermodynamic description of the apparent horizon by changing the effective gravitational coupling and also the relation between the horizon entropy and the horizon area. As a result, the entropy of the horizon is written in terms of an effective gravitational coupling $G_{\mathrm{eff}}$ instead of the usual Newtonian gravitational constant. In this way, the thermodynamic analysis carried out in this work includes the effects of modified gravitational dynamics through the effective coupling as well as through the reconstructed dark-energy sector that controls the cosmological evolution. }

The apparent horizon is chosen since it always exists in FLRW spacetime and satisfies the first law of thermodynamics. In modified gravity theories, the horizon entropy can be expressed in an effective form as \cite{Setare2006, Cai2005}
\begin{equation}
S_h=\frac{A}{4G_{\rm eff}}=\frac{\pi R_A^2}{G_{\rm eff}},
\end{equation}
where $G_{\rm eff}$ denotes the effective gravitational coupling. 

\textcolor{black}{It is important to note that in modified gravity frameworks the gravitational coupling appearing in the horizon entropy is not necessarily identical to the Newtonian gravitational constant. In Hořava–Lifshitz gravity, the modification of the gravitational action and the presence of additional coupling parameters lead to an effective gravitational coupling that can differ from the standard Newton constant. In the infrared limit, where Hořava–Lifshitz gravity is expected to reproduce General Relativity, the effective coupling reduces to the usual gravitational constant. Therefore, in the thermodynamical description of the apparent horizon we employ an effective gravitational coupling $G_{\mathrm{eff}}$, which encapsulates possible deviations from the standard gravitational interaction arising from the modified gravitational dynamics. For the purpose of the present thermodynamical analysis, $G_{\mathrm{eff}}$ is treated as a constant parameter characterizing the gravitational strength at cosmological scales.}

\textcolor{black}{It may be noted that an explicit derivation of the effective gravitational 
coupling from the full Ho\v{r}ava--Lifshitz action is beyond the scope of the present work, and it is treated here as a phenomenological parameter capturing 
the modified gravitational effects.}

Taking the time derivative, we obtain
\begin{equation}
\dot S_h=\frac{2\pi R_A}{G_{\rm eff}}\dot R_A
=\frac{2\pi t}{n^2 G_{\rm eff}}>0,
\end{equation}
indicating that the horizon entropy always increases for an expanding Universe. The associated horizon temperature is \cite{Cai2005}
\begin{equation}
T_h=\frac{1}{2\pi R_A}=\frac{n}{2\pi t}.
\end{equation}
The entropy of the effective cosmic fluid inside the horizon obeys the Gibbs equation \cite{Sheykhi2025,Izquierdo2006}
\begin{equation}
T_h dS_{\rm DE}=d(\rho V)+p\,dV,
\end{equation}
where the volume enclosed by the apparent horizon is
\begin{equation}
V=\frac{4}{3}\pi R_A^3=\frac{4\pi}{3}\frac{t^3}{n^3}.
\end{equation}
Taking the time derivative, we obtain
\begin{equation}
\dot S_{\rm DE}
=\frac{1}{T_h}\left[V\dot\rho+(\rho+p)\dot V\right].
\end{equation}
Assuming the effective conservation equation
\begin{equation}
\dot\rho+3H(\rho+p)=0,
\end{equation}
we find
\begin{equation}
\dot S_{\rm DE}
=\frac{8\pi^2 t^3}{n^4}(1-n)(\rho+p).
\end{equation}
The total entropy of the Universe is defined as
\begin{equation}
S_{\rm tot}=S_h+S_{\rm DE}.
\end{equation}
Its time derivative reads
\begin{equation}
\dot S_{\rm tot}
=\frac{2\pi t}{n^2 G_{\rm eff}}
+\frac{8\pi^2 t^3}{n^4}(1-n)(\rho+p).
\end{equation}

In the present case, from Eqs.~(14) and (15), we can obtain
\begin{equation}
\rho_{\rm DE}+p_{\rm DE}
=\frac{3\lambda-1}{2}\dot{\sigma}^{2}.
\label{rhoplus_simple}
\end{equation}

Substituting Eq.~(\ref{rhoplus_simple}) into the entropy production rate of the dark-energy fluid,
\begin{equation}
\dot S_{\rm DE}
=\frac{8\pi^{2}t^{3}}{n^{4}}(1-n)(\rho_{\rm DE}+p_{\rm DE}),
\end{equation}
we find
\begin{equation}
\dot S_{\rm DE}
=\frac{4\pi^{2}t^{3}}{n^{4}}(1-n)(3\lambda-1)\dot{\sigma}^{2}.
\label{SDE_simple}
\end{equation}

Using Eqs.~(14), (15), and the reconstructed scalar-field kinetic term (31), the entropy production rate of the dark-energy sector reads
\begin{equation}
\dot S_{\rm DE}
=
\frac{96\pi^{2}c^{2}\alpha(2n-1)}{n^{4}}(1-n)\,t
+
\frac{192\pi^{2}c^{2}\beta n^{2}(n-1)}{n^{4}}(1-n)\,\frac{1}{t}.
\label{SDEdot}
\end{equation}

The entropy associated with the apparent horizon is given by
\begin{equation}
\dot S_h=\frac{2\pi t}{n^{2}G_{\rm eff}}.
\label{Shdot}
\end{equation}

Consequently, the total entropy variation of the Universe,
$S_{\rm tot}=S_h+S_{\rm DE}$, is obtained as
\begin{equation}
\dot S_{\rm tot}
=
\frac{2\pi t}{n^{2}G_{\rm eff}}
+
\frac{96\pi^{2}c^{2}\alpha(2n-1)}{n^{4}}(1-n)\,t
+
\frac{192\pi^{2}c^{2}\beta n^{2}(n-1)}{n^{4}}(1-n)\,\frac{1}{t}.
\label{Stotdot}
\end{equation}

\begin{figure}[htbp]
\centering
\includegraphics[width=0.7\linewidth]{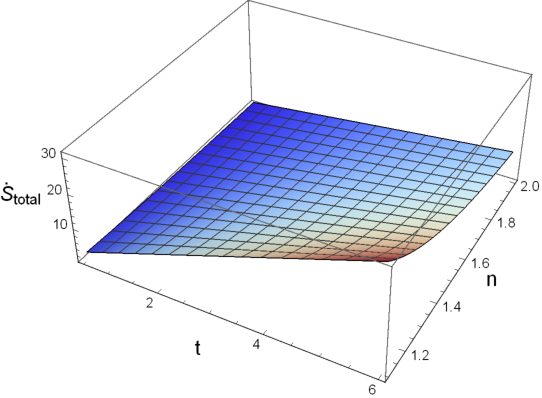}
\caption{
Three-dimensional evolution of the total entropy production rate
$\dot S_{\rm tot}$ as a function of cosmic time $t$ and the power-law
index $n$, for a fixed Gauss--Bonnet coupling parameter $\beta=1$.
The analysis is performed for $1.1 \le n \le 2$, corresponding to an
accelerating universe. The positivity of $\dot S_{\rm tot}$ throughout
the plotted domain confirms the validity of the generalized second law
of thermodynamics.
}
\label{fig:Sdottotal}
\end{figure}

Figure~\ref{fig:Sdottotal} depicts the three-dimensional evolution of the total entropy production rate $\dot S_{\rm tot}$ as a function of cosmic time $t$ and the power-law index $n$ for the Ricci--Gauss--Bonnet dark energy model within Ho\v{r}ava--Lifshitz cosmology as presented in Eq. \eqref{Stotdot}. The monotonic increasing pattern of $\dot S_{\rm tot}$ with $t$ makes it apparent that that there exists dominant contribution of the apparent horizon entropy. The apparent horizon grows continuously with the expansion of the universe, and leads to an irreversible increase in the horizon entropy. This is the source of total entropy production. The plot also shows the dependence of $\dot S_{\rm tot}$ on $n$. For the entire range of $n$, the entropy production rate is positive and there happens to exist a downward tilt as the value of $n$ increases. In the early epochs, the entropy evolution is influenced by the Gauss--Bonnet sector. However, these contributions decay rapidly with cosmic evolution. The non-negative $\dot S_{\rm tot}$ over the entire range of $n$ leads us to conclude that the Ricci--Gauss--Bonnet dark energy model within Ho\v{r}ava--Lifshitz cosmology is thermodynamically consistent and physically viable.

\textcolor{black}{At this juncture, we would like to mention that the thermodynamic description in Ho\v{r}ava--Lifshitz gravity can be different from that in standard Einstein gravity due to the presence of the running parameter $\lambda$. In particular, the entropy associated with homogeneous HL cosmologies can be expressed in terms of the Cardy entropy relation, which explicitly depends on $\lambda$ \cite{Luongo2016}. It is worth mentioning that the thermodynamic treatment adopted in the present work differs from that discussed in Ref.~\cite{Luongo2016}. In Ref.~\cite{Luongo2016}, the entropy of Ho\v{r}ava--Lifshitz cosmology is analyzed from a holographic perspective by expressing the Hamiltonian constraint in terms of the Cardy--Verlinde entropy relation. In this context, it has been shown in \cite{Luongo2016} that the entropy can be written as $S_C=\frac{VH}{2G}\sqrt{\frac{3\lambda-1}{2}}$, indicating that the gravitational entropy acquires a multiplicative correction determined by the HL parameter \cite{Luongo2016}. This result demonstrates that the thermodynamic properties of the HL cosmological framework remain well defined, although they are modified compared with those of standard general relativity. In the present work, the generalized second law of thermodynamics is examined at the apparent horizon by considering the total entropy as the sum of the horizon entropy and the entropy of the cosmic fluid inside the horizon. The influence of HL gravity enters through the modified cosmological dynamics and the effective dark-energy sector reconstructed in this model.}

\textcolor{black}{Thus, the modified gravitational dynamics of Ho\v{r}ava--Lifshitz gravity directly influence the thermodynamic properties of the apparent horizon through the effective gravitational coupling and the modified cosmological evolution.}

\section{Conclusions}
It has been firmly established through the observations of distant Type Ia supernovae that the present Universe is undergoing a phase of accelerated expansion \cite{Novosyadlyj2013, Amendola2010}. This observational evidence suggests the presence of an exotic cosmic component characterized by negative pressure, commonly referred to as dark energy \cite{Novosyadlyj2013}. Such observational outcomes are further supported by independent measurements of the Hubble parameter, large-scale structure, and cosmic matter density \cite{Novosyadlyj2013}. Together, they rule out the possibility of a purely matter-dominated cosmological evolution. Although the simplest explanation is provided by cosmological constant, the theoretical shortcomings of cosmological constant have motivated the cosmologists towards an extensive exploration of dynamical dark energy scenarios and modified gravity frameworks. In this context, constructing viable models capable of reproducing late-time acceleration while remaining consistent with observational and thermodynamical requirements has become a central theme in contemporary cosmology.

In this work, we have investigated a Ricci--Gauss--Bonnet (RGB) dark energy model within the framework of Ho\v{r}ava--Lifshitz cosmology and provided a complete scalar-field reconstruction of the effective dark-energy sector. The study reported here is motivated by the holographic considerations and higher-curvature corrections that arise in modified gravity. The current study has focused on the RGB construction. This approach incorporates both the Ricci scalar and the Gauss--Bonnet invariant. Hence the study deals with local curvature effects as well as higher-order geometric contributions that are expected to be relevant across different cosmological epochs.

We have derived explicit analytical expressions for the effective dark-energy density and pressure (see \eqref{rhoDE} and \eqref{pDE}) under the assumption of a spatially flat FLRW universe with a power-law scale factor. With these expressions we reconstructed the RGB dark energy sector as an effective scalar field within Ho\v{r}ava--Lifshitz cosmology. The behaviour of the reconstructed scalar-field kinetic term and potential throughout cosmic evolution has been analyzed (see \eqref{31} and \eqref{32}). During the evolution, they have not exhibited any divergences for physically admissible parameter choices (see Fig. \ref{fig:dVdsigma} and \ref{fig:Vsigma}). The study presented here reveals a physical picture in which the Gauss--Bonnet contribution dominates the early-time dynamics. On the other hand, the Ricci curvature term governs the late-time accelerated expansion, leading to a smooth transition between different curvature regimes. \textcolor{black}{The present reconstruction explicitly shows that the higher-curvature Gauss–Bonnet contribution controls the early-time dynamics, while the Ricci sector governs the late-time accelerated expansion.}

We examined the cosmological dynamics of the model by observing the evolution of the effective equation-of-state parameter (see \eqref{eq:wDE_full}) reconstructed accordingly. Based on the choice of the model parameters, the RGB dark energy has been found to show a quintessence-like behavior and it has been further observed to remain close to, or deviate from, the cosmological-constant boundary without encountering instabilities (see Fig.~\ref{fig:wDE}). Both the scalar-field kinetic term and reconstructed potential have been found to show explicit time dependence that prevents the system from settling into a pure vacuum-dominated phase. Hence, a natural realization of a dynamical dark-energy scenario within the Ho\v{r}ava--Lifshitz framework is observed.

We further analyzed the classical stability of the model by obtaining the squared speed of sound of the effective dark-energy fluid. The results indicate that classical stability can be achieved within a physically motivated region of parameter space, particularly when the Gauss--Bonnet contribution dominates over the Ricci term (see Fig.~\ref{fig:vs2_3D}). This highlights the crucial role of higher-curvature effects in ensuring the viability of dark-energy models formulated in Ho\v{r}ava--Lifshitz cosmology.

An important aspect of the present work was the examination of thermodynamical consistency. By considering the apparent horizon as the thermodynamically relevant boundary and computing the entropy evolution of both the horizon and the cosmic fluid, we tested the validity of the generalized second law of thermodynamics. We found that the total entropy production rate (see \eqref{Stotdot}) remains non-negative throughout cosmic evolution for a wide range of model parameters, confirming that the RGB dark energy model within Ho\v{r}ava--Lifshitz cosmology is thermodynamically consistent. The monotonic increase of the total entropy is primarily driven by the growth of the apparent horizon, while the contributions from the dark-energy sector decay at late times (Fig.~\ref{fig:Sdottotal}).

In summary, the Ricci--Gauss--Bonnet dark energy model embedded in Ho\v{r}ava--Lifshitz cosmology provides a dynamically rich, classically stable, and thermodynamically viable description of late-time cosmic acceleration. The framework successfully combines higher-curvature geometric effects with anisotropic scaling, offering a unified and consistent approach to dark-energy dynamics beyond the standard cosmological constant. Future studies may focus on confronting the model with observational data and performing a detailed perturbation analysis to further assess its phenomenological viability.

While concluding let us comment on the outcomes reported here from some relevant existing studies in the literature. In the context of standard FLRW cosmology, \cite{Saridakis2018,Iqbal2018,Ahmed2020} formulated Ricci--Gauss--Bonnet dark energy primarily as a holographic or effective fluid model. While these works successfully demonstrate late-time acceleration and, in some cases, thermodynamical consistency, the dark energy sector remains phenomenological. The present work, contrary to the existing literatures,  has worked on a scalar-field reconstruction approach for the RGB dark energy sector within Ho\v{r}ava--Lifshitz framework, where the work identified explicit kinetic and potential terms responsible for the cosmic dynamics. An earlier notable work on Gauss--Bonnet dark energy in Ho\v{r}ava--Lifshitz gravity focused on effective fluid descriptions and horizon thermodynamics \cite{Lepe2018,Sheykhi2007,Sheykhi2009,Cai2010HL}. The current study, on the other hand, addressed the reconstructed field dynamics and  demonstrated a smooth and intrinsic transition from a Gauss--Bonnet--dominated early-time regime to a Ricci-driven late-time accelerated phase as a consequence of the reconstructed scalar dynamics. 

As far as thermodynamics is concerned, existing studies have successfully shown the sensitivity to model assumptions in different dark energy and modified gravity frameworks \cite{Izquierdo2006,Sheykhi2025}. More recently, generalized entropy-based cosmologies have demonstrated consistency between horizon thermodynamics and modified Friedmann equations under specific conditions \cite{Luciano2023}. In that context, in the current work, we find that the validity of the generalized second law in the Ricci--Gauss--Bonnet model within Ho\v{r}ava--Lifshitz cosmology is robust over a wide range of parameters, with the apparent horizon entropy providing the dominant contribution to the total entropy budget. Overall, the results obtained here  demonstrated the inclusion of higher-curvature corrections and obtained a dynamically viable and thermodynamically stable realization of Ricci--Gauss--Bonnet dark energy in the Ho\v{r}ava--Lifshitz framework. \textcolor{black}{It is useful to compare the present results with earlier works on thermodynamics in Ho\v{r}ava--Lifshitz cosmology. For example, Luongo \textit{et al.} \cite{Luongo2016} studied the entropy structure of HL cosmology using the Cardy--Verlinde relation and showed that the entropy depends on the running parameter $\lambda$. Their work  focused on the holographic description of entropy in HL gravity. On the contrary, the present work studies the validity of the generalized second law of thermodynamics for the reconstructed Ricci--Gauss--Bonnet dark energy model in Ho\v{r}ava--Lifshitz cosmology. Dunsby et al. \cite{Dunsby2024} investigated a generalized unified dark energy model where a tachyonic fluid interacts nonminimally with an additional scalar field possessing vacuum energy arising from a symmetry-breaking mechanism. They \cite{Dunsby2024} had shown that after the symmetry-breaking transition, the effective equation of state behaves as a mixture of a generalized Chaplygin gas (GCG). The results obtained in the present study show that the total entropy does not decrease at the apparent horizon for the considered ranges of parameters. This indicates that the model remains thermodynamically consistent.}

\textcolor{black}{It may be noted that the thermodynamic treatment adopted in the present work differs from that discussed in Ref.~\cite{Luongo2016}. While Ref.~\cite{Luongo2016} derives the entropy within a holographic framework based on the Cardy--Verlinde relation, the present study considers an effective thermodynamic description at the apparent horizon, where the effects of Ho\v{r}ava--Lifshitz gravity are taken into account through the modified cosmological dynamics and the reconstructed dark-energy sector.}

Future studies may test the present model with observational data and may also perform a more detailed perturbation analysis within the Ho\v{r}ava--Lifshitz framework in order to further examine the physical viability of the model. It should be mentioned that the stability analysis carried out in this work is based on the effective sound-speed criterion, which gives a first indication of the classical stability of the reconstructed dark-energy sector. A full perturbation analysis including scalar perturbations and possible stability issues such as ghost and gradient instabilities is beyond the scope of the present work and may be considered in future studies. In conclusion, it may be noted that the study reported in this paper demonstrates the possibility of reconstructing a stable RGB dark energy sector within the Hořava--Lifshitz cosmological framework. The viability of the model has been illustrated by considering a power-law form of the scale factor. Nevertheless, a more comprehensive understanding of the dynamical behaviour of the model would require a detailed dynamical system analysis. Such an investigation may help to ascertain whether the obtained stable solutions can act as global attractors even in the absence of an imposed background expansion. \textcolor{black}{Future directions will focus on the possibility of producing particles in contexts such as the one here developed in \cite{future1,future2}.}

\section*{Acknowledgment}
The author sincerely thanks the anonymous reviewers for their insightful suggestions. The author gratefully acknowledges the hospitality and support of the Inter-University Centre for Astronomy and Astrophysics (IUCAA), Pune, where part of this work was carried out during a scientific visit in December 2025. The author also sincerely acknowledges Souptik Chattopadhyay for his assistance in preparing the final version of the manuscript.

\section*{Declaration on the use of Generative AI}

The author declares that generative artificial intelligence tools were used solely for language editing and improvement of clarity and presentation of the manuscript. The author is responsible for the scientific content of the paper.


\end{document}